\documentclass[sigconf]{acmart}
\usepackage{colortbl}
\usepackage{graphicx}
\usepackage{threeparttable}
\usepackage{amsthm}
\usepackage{multirow}
\usepackage{supertabular}
\usepackage[figuresright]{rotating}
\usepackage{tabularx}
\usepackage{flushend}
\usepackage[labelfont=bf]{caption}
\AtBeginDocument{%
  \providecommand\BibTeX{{%
    \normalfont B\kern-0.5em{\scshape i\kern-0.25em b}\kern-0.8em\TeX}}}

\copyrightyear{2022}
\acmYear{2022}
\setcopyright{acmcopyright}\acmConference[ICSE-SEIP '22]{44nd International Conference on Software Engineering: Software Engineering in Practice}{May 21--29, 2022}{Pittsburgh, PA, USA}
\acmBooktitle{44nd International Conference on Software Engineering: Software Engineering in Practice (ICSE-SEIP '22), May 21--29, 2022, Pittsburgh, PA, USA}
\acmPrice{15.00}
\acmDOI{10.1145/3510457.3513054}
\acmISBN{978-1-4503-9226-6/22/05}
\begin{document}

\title{A Cross-Company Ethnographic Study on Software Teams for DevOps and Microservices: Organization, Benefits, and Issues}

\author{Xin Zhou$^*$, Huang Huang$^+$, He Zhang$^*$, Xin Huang$^*$, Dong Shao$^*$, Chenxin Zhong$^*$}
\affiliation{%
 \institution{$^*$State Key Laboratory of Novel Software Technology, Software Institute, Nanjing University, Nanjing, Jiangsu, China}
%\diamondsuit
\institution{$^+$State Grid Nanjing Power Supply Company, Nanjing, Jiangsu, China}
%  \city{Nanjing}
%  \state{Jiangsu}
%  \country{China}
}
\email{{njuxinzhou, sgcc.huang.huang}@gmail.com, hezhang@nju.edu.cn}
\email{njuhuangx@outlook.com, dongshao@nju.edu.cn, zhongcx@smail.nju.edu.cn}

\renewcommand{\shortauthors}{Xin Zhou et al.}
\renewcommand{\shorttitle}{Software Teams for DevOps and Microservices}
%
% The "author" command and its associated commands are used to define the authors and their affiliations.
% Of note is the shared affiliation of the first two authors, and the "authornote" and "authornotemark" commands
% used to denote shared contribution to the research.

%
% By default, the full list of authors will be used in the page headers. Often, this list is too long, and will overlap
% other information printed in the page headers. This command allows the author to define a more concise list
% of authors' names for this purpose.

%
% The abstract is a short summary of the work to be presented in the article.
\begin{abstract}
\textit{\textbf{Context:}} DevOps and microservices are acknowledged to be important new paradigms to tackle contemporary software demands and provide capabilities for rapid and reliable software development. Industrial reports show that they are quickly adopted together in massive software companies. However, because of the technical and organizational requirements, many difficulties against efficient implementation of the both emerge in real software teams.
\textit{\textbf{Objectives:}}
This study aims to discover the organization, benefits and issues of software teams using DevOps \& microservices from an immersive perspective.
%From an immersive perspective, this study aims to discover the reciprocal impacts between DevOps \& microservices and the software teams who are implementing them.
\textit{\textbf{Method:}} An ethnographic study was carried out in three companies with different business, size, products, customers, and degree of globalization. All the three companies claimed their adoption of DevOps and microservices. Seven months (cumulative) of participant observations and nine interviews with practitioners were conducted to collect the data of software teams related to DevOps and microservices. A cross-company empirical investigation using grounded theory was done by analyzing the archive data. 
\textit{\textbf{Results:}} The virtual software teams were organized for adopting DevOps and microservices under the stubborn organizational structure. The adoption of DevOps and microservices brings benefits to rapid delivery, ability improvements and burden reduction, whilst the high cost and lack of practical guidance were emerged. Two major issues of adopting DevOps and microservices in software teams (i.e. fragmentary DevOps and abuse of microservices) were found 
common in the companies. Moreover, our observations and interviews reflect that in software teams, the relationship between DevOps and microservices is not significant, which differs from the relationship described in the previous studies. Four lessons for practitioners and four implications for researchers were discussed based on our findings.
% Although DevOps has brought many benefits to rapid delivery, the adoption of DevOps seems fragmentary in reality, which might be the opportunity for studying DevOps maturity. The pursuit of excessive agility and blind conformity may lead to the abuse of microservices, which further results in some pains in practice. One root cause of the problems associated with both DevOps and microservices is the stubborn organizational structure, in which culture change remains a longer way to go than the techniques.
% In addition, as a result of developers' work limitations, the lack of training and insufficient knowledge of opinion leaders, the correlation between DevOps and microservices is not significant in software teams. Eight lessons are further distilled for practitioners and researchers to implement and study DevOps \& microservices.
\textit{\textbf{Conclusion:}} Our findings contribute to the understanding of the organization, benefits and issues of adopting DevOps and microservices from an immersive perspective of software teams. 
% concepts, practices, and perceptibility of DevOps and microservices from the immersive perspective of software teams.
\end{abstract}


%
% Keywords. The author(s) should pick words that accurately describe the work being
% presented. Separate the keywords with commas.



%
% A "teaser" image appears between the author and affiliation information and the body 
% of the document, and typically spans the page.

%
% This command processes the author and affiliation and title information and builds
% the first part of the formatted document.
\maketitle
%%%\vspace*{-1.0ex}
\section{Introduction}
With the diverse and quickly changing user demands and the even fierce market competition, software organizations have to constantly improve the development process to deliver value quickly and implement more complex software systems. To address the problems arisen in the development process, DevOps, which was proposed from industry in 2009~\cite{allspaw200910+} and originated from the combination of \textit{``development''} and \textit{``operations''}, offers a series of controllable and actionable Software Engineering (SE) strategies that can extend software design changes~\cite{artac2017devops}. Moreover, DevOps emphasizes the emotional resonance and cross-stream collaboration within or between development teams and operation teams as well~\cite{dyck2015towards}. 

Emerging from Agile developer communities, microservices~\cite{fowler2014micro}, which is the portmanteau of \textit{``micro''} and \textit{``services''}, was considered as the architecture solutions for continuous architecture setting in DevOps~\cite{balalaie2016microservices,taibi2018continuous, chen2018microservices}. Microservices achieve benefits (e.g., on maintainability, reusability, scalability, availability and automated deployment) by decomposing the monolith into small services that communicate with each other through light weight mechanisms~\cite{li2019dataflow}.

\begin{figure}[!htbp]
  \begin{center}
  \vspace*{-2.0ex}
    \includegraphics[width=0.4\textwidth]{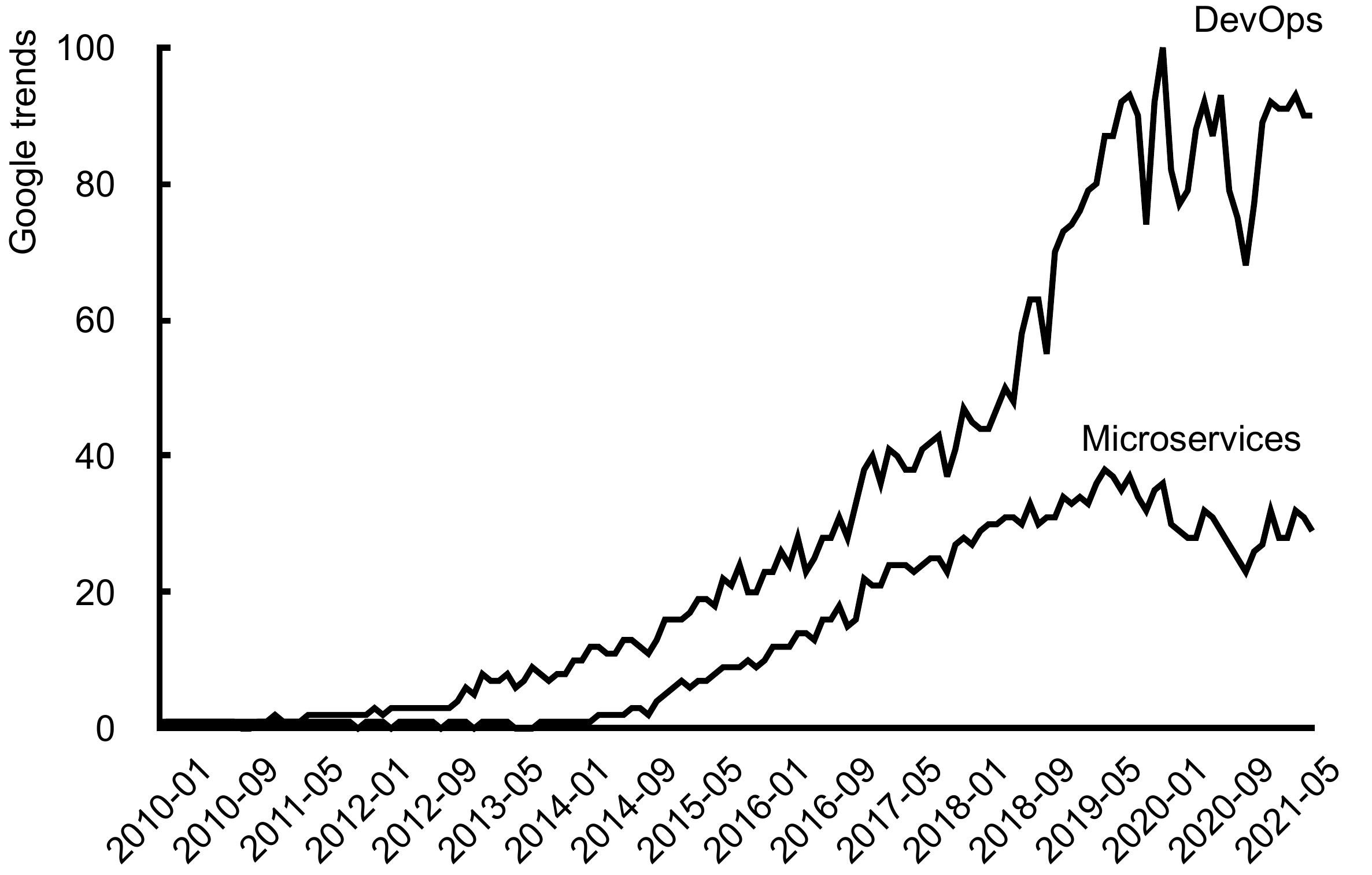}
    \vspace*{-2.0ex}
  \end{center}
    \caption{Google trends on DevOps and microservices} 
    \label{fig:trends}
    \vspace*{-2.0ex}
\end{figure}

Both DevOps and microservices have received increasing attention from researchers and practitioners over the past decade~\cite{waseem2020systematic}. Figure~\ref{fig:trends} shows Google search trends (by topic) for the two concepts since 2010~\footnote{Accessed September 1, 2021.}, with a high degree of consistency. 
They are commonly adopted and implemented together in many software companies. Puppet's report on DevOps~\cite{brow20} pointed out that microservices correlate with high performance under DevOps contexts while the report on microservices from O'Reilly~\cite{ford20} shows that more than half the companies with microservices utilize continuous deployment, the highest bar for automated DevOps.
%After a period of sustained increase in search popularity, they have been leveling off and decreasing slightly since 2019. Although this fluctuation might be due to the saturation of concept evolution or some force majeure factors, the adoption of these two concepts to companies is also worth considering. 

%Some IT giant companies are promoting their solutions for DevOps and microservices, e.g., Amazon Web Services~\cite{rich18}, Microsoft Azure~\cite{microsoftazure} and Google Cloud~\cite{googlecloud}. Researchers have also begun to move their attention to DevOps and microservices in industry~\cite{lwakatare2019devops, fritzsch2019microservices}. They reflected on the use of DevOps and microservices in actual industrial scenarios to help researchers and practitioners improve their understanding of the concepts and practices, thereby promoting the further evolution of DevOps and microservices. These studies focused on technical improvements or overall characteristics of DevOps and microservices. Although they studied the DevOps and microservices practices in industry, they paid more attention on the understandings at organizational level, whilst lack the immersive thinking from software developers and teams, who are the atomic units of software development activities. 

Some IT giant companies are promoting their solutions for DevOps and microservices, e.g., Amazon Web Services, Microsoft Azure and Google Cloud. Researchers have also begun to move their attention to DevOps and microservices in industry~\cite{lwakatare2019devops, fritzsch2019microservices}. They reflected on the use of DevOps and microservices in actual industrial scenarios to help researchers and practitioners improve their understanding of the concepts and practices, thereby promoting the further evolution of DevOps and microservices. These studies focused on technical improvements or overall characteristics of DevOps and microservices. Although they studied the DevOps and microservices practices in industry, they paid more attention to the understanding at organizational level, whilst lacking the immersive thinking from software developers and teams, who are the atomic units of software development activities.

%On the other hand, SE is a socio-technical endeavor where stakeholders should benefit from the contributions on technical aspects~\cite{storey2020software}. However, few studies use strategies that focus on human and social aspects, which implies the complex, socio-technical context of SE problems in real practice scenarios would be missed in research~\cite{storey2020software}.

The objective of this ethnographic study is to provide an immersive perspective, revisiting software development activities as a member of software teams, to observe the adoption of DevOps and microservices in software teams and to discover the organization, benefits, and issues of software teams that adopt DevOps and microservices.
% impacts of DevOps and microservices on software teams as well as the feedback from software teams on DevOps and microservices. 
The focus on software team's daily work could offer an insight into the specific form of DevOps and microservices in real software development, which might be different from the original vision of the software organization, and also contribute to the practical implementation of DevOps and microservices. Specifically, activities of the teamwork from the three companies were observed in participant's view as well as a series of open-ended interviews with practitioners in these companies were conducted to drive the investigation of the reciprocal impacts between software teams and the two concepts. 

This study contributes to the understanding of software teams for DevOps and microservices by providing an immersive perspective. Organizational structure has become a barrier to the further evolution of DevOps and microservices in software organizations where strong organizational support and cultural accumulation are required. Under this circumstance, virtual software teams were organized for the adoption of DevOps and microservices. Rapid iteration and deployment, enhanced ability and reduced burden are found as the three main benefits in software teams, whilst high cost and lack of practical guidance were the two major pains. The adoption and implementation of DevOps and microservices in reality were found fallen short of the expectations of software organizations. Specifically, the departmental barriers that still exist in many software teams while the public service in the product expands over the course of business development. Moreover, the relationship between DevOps and microservices in software teams was found with weak significance, which reminds researchers and advocates of more attention to the real situation of practices in daily work. 
% More training and stronger upper-level supports are essential for the success of DevOps and microservices.

% The rest of this paper is organized as follows. Section~\ref{sec:background} introduces DevOps and microservices followed by the details about the steps of conducting this ethnographic study in Section~\ref{sec:method}. Section~\ref{sec:result} describes the findings and discussions from DevOps, microservices and team organizations. Section~\ref{sec:validity} discusses the internal and external threats to validity of this paper. Section~\ref{sec:conclusion} draws the conclusions.

%%%\vspace*{-1.0ex}
\section{Background and Related Work}
\label{sec:background}
This section provides an overview of DevOps and microservices as well as some existing empirical studies focusing on their adoption and implementation in practice.

\subsection{DevOps and Microservices in Industry}
Despite both DevOps and microservices originated in the industry, there have been a number of related studies reported in academia, industry-orientation should be the original intention of the research on these two concepts.

Lwakatare et al.~\cite{lwakatare2019devops} conducted a multiple-case study in five companies with successful DevOps implementations, enhanced the definition of DevOps by emphasizing automation practices and confirmed the capabilities of DevOps for software development teams to deploy software in production environment fast. In addition to the technical practices, a supportive culture and mind-set, especially from senior management and customers that could effect cross-stream collaboration, was identified as critical in the long-term activity of the adoption and implementation of DevOps as well. 
Using grounded theory approach, Luz et al.~\cite{luz2019adopting} generated a theory to explain the successful DevOps adoption in fifteen companies and proposed a DevOps workflow, which was evaluated at a Brazilian Government institution. They found that the existing automation and tools can be enough for DevOps whilst collaboration is the core DevOps concern.

\begin{figure*}[!htbp]
  \begin{center}
   \vspace*{-2.0ex}
    \includegraphics[width=0.94\textwidth]{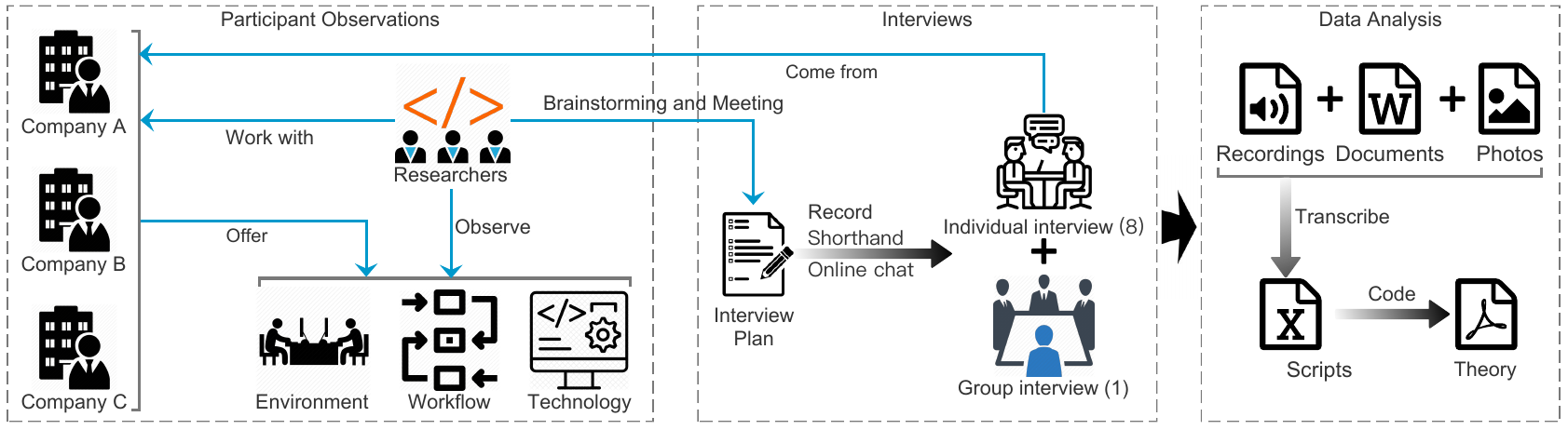}
    \vspace*{-2.0ex}
  \end{center}
    \caption{Research process of this ethnographic study} 
    \label{fig:process}
    \vspace*{-2.0ex}
\end{figure*}

Zhang et al.~\cite{zhang2019microservice} carried out a series of industrial interviews in thirteen companies to study the gaps between the visions and the reality as well as the benefits and sufferings of microservices around the nine characteristics claimed by Fowler and Lewis~\cite{fowler2014micro}. They claimed that organizational transformation, which is associated to the pains deserves further research directions. Bogner et al.~\cite{bogner2019microservices} analyzed fourteen systems from ten companies with seventeen interviews, where most systems did not show the high degree of technological diversity as commonly expected with microservices. As the number of coarse-grained services continues to increase, the boundary line between service- and microservices-based systems is blurring.

%The concept of microservices appears commonly in industrial studies on DevOps, while DevOps is also involved in the studies on microservices. 
Several studies focused on both DevOps and Microservices in industry settings.
For example, Rademacher et al.~\cite{rademacher2020deriving} conducted two case studies to validate the applicability of the model-driven methodology, which uses viewpoint-based microservices modeling languages and respects DevOps teams' concerns. The industrial case study conducted by Shahin et al.~\cite{shahin2020role} showed that the characteristics of microservices meet the keys to DevOps success.

These empirical studies focus more on technical improvements to DevOps and microservices, and are more concerned with the overall characterization of DevOps and microservices in industry. They provide objective and rational descriptions of DevOps and microservices from a non-participant perspective. Although a number of interviews with practitioners and some observations on the practices were conducted in these studies, they are mostly carried with the researchers' prior experience, thus ignoring some small but key influencing factors (e.g., personal experience at each software phase) in practice, especially from the human and social aspects. This ethnographic study increases the possibility of meeting important moments in the research process through the actual participation of researchers in the software teams, the basic unit of software development, to conduct participant observations and interviews. The interaction between software teams and DevOps together with microservices exposed in these processes could also contribute to the method improvement of DevOps and microservices~\cite{sharp2016role}.

\subsection{Empirical Research on Software Teams}
Recent research on software teams mainly focuses on those for agile methods, which emphasize the importance of teamwork~\cite{nmm05}. High-quality teamwork can contribute to the success of projects in agile development, where the team leads and members have significant influences on the teamwork quality and performance (different on small and large projects)~\cite{lbds18}. Team turnover of agile software teams was also studied. To better understand the change in the agile team, Hilton and Begal~\cite{hb18} conducted interviews and surveys in some software organizations with high turnover rates to explore the reasons why members left and joined. They further analyzed six-year data from employee database to distill the costs and benefits of the team’s turnover~\cite{hb18}.
Researchers and practitioners were committed to exploring the organization of agile teams as well. Kniberg et al. \cite{kniberg2012scaling} first introduced Spotify Model, a software development model driven by several autonomous squads. Salameh et al.~\cite{salameh2018influential, salameh2019spotify} used a variety of methods (e.g., case study and interview) to conduct a series of studies on the organization of this agile development team. Zhang et al.~\cite{zhang2020fireteam} analyzed the organization of fireteam, one agile development team, via their four key activities in software development. Similarly, this study focuses on agile software teams using DevOps \& microservices to explore their organization, benefits, and issues from an immersive perspective.

%%%\vspace*{-1.0ex}
\section{Research Methodology}
\label{sec:method}
This section describes the research methodology step by step in this study. 
Following the guidelines~\cite{zhang2019ethnographic}, we conducted an ethnographic inquiry, which is used to study people and cultures~\cite{sharp2016role}. Moreover, an ethnographic study offers insights into relationships between humans, process, technology and environment~\cite{zhang2019ethnographic}. Aiming to discover particular groups of people's perspectives in an organization~\cite{johannesson2012design}, researchers endeavor to understand DevOps and microservices in software teams by investigating the daily reality of individuals or teams in various environments (e.g., the working space, communications, meetings and develop environments). The value of ethnography lies in its basis on pure observation, which is empirical in nature, and it is used to develop theories from an overall explanation~\cite{bibri2019awareness}.
Figure~\ref{fig:process} shows the process of this study. 
%and the study design, data collection and data analysis are future detailed in the following subsections.

\begin{table*}[!htbp]
    \vspace*{-2.0ex}
	\caption{Profiles of and interviewees from software organizations}
	\label{table:company}
	\vspace*{-2.0ex}
	\scriptsize
	\begin{center}
		\begin{tabular}{p{15mm}p{55mm}p{45mm}p{40mm}}
			\hline
			\rowcolor{gray!30}
			Information & Company 1 (C1) & Company 2 (C2) & Company 3 (C3)\\
			\hline
			Domain & Information technology \& Telecommunication & Big data \& Artificial intelligence & Software development\\
			Size & 1000+ developers & 100-1000 developers & 50-100 developers\\
			Products & Infrastructure for information and communication technology & Databases and cloud platforms & Work management system\\
			Customers & Government and public utilities, colleges, financial institutions, energy companies, transportation companies, individual users & Government and public utilities, colleges, financial institutions, transportation companies & Colleges, catering companies, individual users\\
			Globalization & Global enterprise & Domestic company, starting to go global & Local company\\
			Experiences & DevOps (5 years), Microservices (6 years) & DevOps (1 year), Microservices (5 years) & DevOps (3 years), Microservices (5 years) \\
			Interviewees & Developer, software architect(4) & Developer, (assistant) project manager(2) & Developer/architect, product owner, CTO\\
			\hline
		\end{tabular}
	\end{center}
	\vspace*{-3.0ex}
\end{table*}

\begin{table*}[!htbp]
	\caption{Observation items in the three companies}
	\label{table:observation_items}
	\vspace*{-2.0ex}
	\scriptsize
	\begin{center}
		\begin{tabular}{p{28mm}cccp{125mm}}
			\hline
			\rowcolor{gray!30}
			Observation items & C1 & C2 & C3 & Description\\
			\hline
			Company culture & $\surd$ & $\surd$ & $\surd$ & \textit{Company culture} provides a great help for the investigators to familiarize themselves with the company and understand the reasons for some decisions. The history, brochures, training courses for new employees, and employees’ sense of identity with the company are focused on. \\
			Working environment & $\surd$ & $\surd$ & $\surd$ & \textit{Working environment} includes the working time, working spaces, afternoon tea and employee care in the company. Competitive products and the company's living environment will also be considered in this observation item. \\
			Communications & $\surd$ & $\surd$ & $\surd$ & \textit{Communication} mainly appears between the investigators and other employees. Communication tools, contents, frequency, and efficiency are all concerned when observing. \\
			Ongoing development processes & $\surd$ & $\surd$ & $\surd$ & \textit{Ongoing development processes} focused on the processes used in the software teams and they are commonly specified in the documents. Documents and engineer behavior are considered in conjunction with the observation. In addition, the researchers will actually participate in the processes. \\
			Development environment & $\times$ & $\surd$ & $\surd$ & \textit{Development environment}, differs from the processes, pays more attention to some details such as language and tools. Some development resources (e.g., requirement documents, UI design drawings, source codes, operation reports) are also included in this observation items. \\
			Meetings & $\times$ & $\surd$ & $\surd$ & \textit{Meetings} includes stand-up meetings, weekly meetings, business meetings and some other meetings. The topics discussed in the meeting, the organization format and the performance of the meeting participants will be observed. \\
			\hline
			
		\end{tabular}
	\end{center}
	\vspace*{-3.0ex}
\end{table*}

%%%\vspace*{-1.0ex}
\subsection{Research Questions}
%The multi-case study gives the possibility to investigate the practices in real context~\cite{yin2009case, runeson2009guidelines, runeson2012case}. 
This study aims to explore the organization, benefits, and issues of software teams using DevOps \& microservices in industrial contexts. With this objective, the following Research Questions (RQs) are formulated to guide this ethnographic study:
\begin{description}
	\item[RQ1.] \emph{How do companies organize software teams for DevOps and microservices?}
	\item[RQ2.] \emph{What are the pains and gains of DevOps and microservices to the teamwork?}
	\item[RQ3.] \emph{What issues do these software teams have when adopting DevOps and microservices?}
\end{description}

% To be specific, this RQ aims at discovering the impacts of DevOps and microservices on software teams' daily work as well as the impacts of software team organizations (e.g., the composition of team members, team size, team membership) on the adoption and implementation of DevOps and microservices. 
%, focusing more on the individual aspect than the organizational aspect. RQ2 focuses on 
%The two RQs form a conceptual loop from software teamwork towards DevOps and microservices.

%%%\vspace*{-2.0ex}
\subsection{Access to Ethnographic Fields}

This research project began in January 2020 with a total of seven months of participant observations in the three companies at different scales.
During the observations, the principal investigator (who took a leading role in this ethnographic study, including introducing the methods of observation and designing original semi-structured questionnaires for interviews) and the other authors conducted the participant observations in each company. This is a common ethnographic approach to immerse one or more researchers in the natural environment of the companies~\cite{creswell2017research}. Table~\ref{table:company} shows the basic information about the three companies. They differ in business (domain), size, product, customer, and degree of globalization. All the three companies claimed to use DevOps and microservices. Therefore, they offer similar environments (DevOps and microservices) in different states of corporation development, which allows comprehensive information to be observed in the period of this study. This work follows the checklist proposed in~\cite{zhang2019ethnographic}: multiple examples should be used while the individual field study is short in duration (far less than 8 months).

Two research students from the authorship applied for internships online or through company-college partnerships that offered them opportunities to join the teams and access the real contexts. 
After that, with the industry mentors' help, the researchers understand the company culture and software team working process through various types of documents.
During the internships, the researchers observed the practice status quo in processes of teamwork, which is elaborated in Section~\ref{sec:data_collection}. %The engineers and development processes in researchers' teams are focused on. 
They joined the software teams as team members and participated in their engineering activities, thus the research plan would not disturb the daily work of any other members.

%%%\vspace*{-1.0ex}
\subsection{Data Collection}
\label{sec:data_collection}
Ethnography involves a variety of data collection techniques~\cite{hm07,fm10}, where the field study is the major method~\cite{st99}, and participant observations are most used in ethnographic field study in SE~\cite{zhang2019ethnographic}. Interview is another important means to gathering data in ethnographic studies~\cite{hugh95,hm17}. Therefore, this ethnographic study combines the participant observations and interviews to achieve the triangulation of data collection, which is the basis of one ethnographic study and could help secure the quality of data and the validity of findings~\cite{fm10}.

%%%\vspace*{-1.0ex}
\subsubsection{Participant observations}

The participant observations were conducted in the three companies into three time periods between January and December 2020. First, the principal investigator (one of the research students) served as an intern product owner, which is a part-time job, in C3 from January to April 2020. Note that due to the outbreak of the Covid-19 pandemic, the company promoted working from home for about three months and the observation was carried out online during most of this period. Because of the assistance of remote collaborative office and meeting tools, it did not affect the observation. After that, the principal investigator was employed by C2 as a full-time assistant project manager from July to August. In December 2020, the principal investigator and the other research students joined C1 as the contributors of the company-college collaboration project to observe the practices related to DevOps and microservices. During the participant observations, the investigators worked with software teams, observed their daily activities as well as the behaviors of the team members and other employees, and shared insider’s views of what was happening in these organizations. 

In general, the principal investigator and the other authors observed the DevOps and microservices related practices in these companies from three aspects of environment, workflow and technology. Table~\ref{table:observation_items} shows some specific observation items in this study and the symbol `$\surd$' stands for the observation item being observed in the company.

Note that because the research students got access into C1 through the company-college collaboration project, the observation of their environment was largely limited to the analysis of source code but the participation in their internal meetings was not allowed. This has caused some adverse effects (e.g., misunderstanding of some jokes) on our observations. Hence, a follow-up group interview (cf. Section~\ref{sec:interview}) was conducted to observe the relationships among employees in C1.

%%%\vspace*{-1.0ex}
\subsubsection{Interviews}
\label{sec:interview}
The interviews in C2 and C3 were conducted after the principal investigator left the companies, and the interviewees were selected from the team members whom the principal investigator ever worked with. The interviews in C1 were conducted together with the participant observation, and the interviewees in Table~\ref{table:company} were invited from the software teams collaborated in the college projects. 
%Table~\ref{table:interviews} shows interviewees of the nine interviews in three companies. 
Note that a group interview was carried out with a developer and three software architects from different teams (not limited to the collaborating team). The developers in C1 were interviewed twice. The developer interviewed in C3 is also responsible for some design decision-making at architecture level. Because of the small size of C3, the CTO is also involved in the daily development work as a team leader.

% \begin{table}[!htbp]
%     %%%\vspace*{-1.0ex}%
% 	\caption{The interviews conducted with practitioners}
% 	\label{table:interviews}
% 	%%%\vspace*{-3.0ex}
% 	\scriptsize
% 	\begin{center}
% 		\begin{tabular}{cl}
% 			\hline
% 			\rowcolor{gray!30}
% 			Company & Interviewee(s)\\
% 			\hline
% 			C1 & Developer, software architect(4) \\
% 			C2 & Developer, (assistant) project manager(2)\\
% 			C3 & Developer/architect, product owner, CTO\\
% 			\hline
% 		\end{tabular}
% 	\end{center}
% 	%%%\vspace*{-2.0ex}
% 	* This group interview included developer and architects from different teams
% 	** This developer is also responsible for some of the architecture work in C3
% \end{table}

Driven by the RQs, the interview plan was jointly developed by all authors through brainstorming and meetings. For the setting of interview questions, we followed the principle that interviewees should elaborate more on their own views. As a result, a four-topic interview plan emerged:

\begin{description}
	\item[Topic1.] \emph{Job description and main responsibilities of interviewees.}
	\item[Topic2.] \emph{The organization of teamwork.}
	\item[Topic3.] \emph{Interviewees' stories with DevOps.}
	\item[Topic4.] \emph{Interviewees' stories with microservices.}
\end{description}

Topic1 in C2 and C3 is used to find out whether the interviewees' responsibilities ever changed since the principal investigator left the company. In C1, this topic is helpful because the interviewees came from different teams. Topic2 discusses team organization and includes the impacts of their adopted practices on nontechnical issues such as organizational structure and membership. Topic3 and Topic4 contain the interviewees' motivations and journey to learn DevOps and microservices as well as their daily work on related the related content. 

The discussion about these topics in interviews is open. ``Let the interviewee say more'' is our principle of conducting interviews. Because the interviewees are not in the exact same context, each interview might contain some unique topic-specific questions. For example, we asked what DevOps practices were used during the \textit{requirements analysis} for project managers and product owners. These topic-specific questions were limited to the scope of software teamwork and the interviewees' own experiences on DevOps and microservices, which could help us harvest immersive results from a comfortable interview atmosphere.

%%%\vspace*{-1.0ex}
\subsection{Data Analysis}
Straussian Grounded Theory (GT)~\cite{stol2016grounded, strauss1998basics}, one of the methods to build theories and construct the reality in SE through interactions with language and communication~\cite{sjoberg2008building}, was employed as the data analysis technique in this ethnographic study since it provides a systematic method for generating conceptual theories from qualitative data sources~\cite{kroeger2014understanding}. Not starting from a preconceived conceptual framework, GT generates theories that evolve with the research process through the continuous interactions between the collected data and data analysis~\cite{glaser2009discovery,adolph2011using,stol2016grounded}. 
%Hence, GT is considered as ``methodologically dynamic''~\cite{glaser2009discovery,ralph2015methodological}. In GT, issues about fit, relevance, workability, and modifiability are more important than internal validity~\cite{glaser1998doing,glaser2009discovery}. 
%The theory from GT is based on qualitative analysis, and even if there are probabilistic relationships between concepts, such statistical findings have no basis in grounded theory~\cite{glaser1998doing}.

Open coding, axial coding and selective coding are the three major strategies in GT. %A description of these strategies using in this ethnographic study is as below.
Open coding is to identify the conceptual categories from the original data, which is the first step of GT. The code generated in this step is used to retain the exact words of the interviewees and the participant observations as much as possible. For instance, combined with the observed software development workflow (Figure~\ref{fig:SIMU}), ``\textit{in some internal product feedback online chat groups, software developers will also raise their discomfort when using the product, and will ask the product owner to improve the product for them}'' from C3 was labeled as ``\textit{online group communication in C3}'' and ``\textit{product feedback is not integrated in the pipeline in C3}''. The observations on the ongoing development processes in C2 shows that the requirements management systems (JIRA) and code repositories (GitLab) are linked in the process, but software developers do not actually associate requirements with their own code in daily work. This was labeled as ``\textit{separation of requirements and development in daily work in C2}''.

Axial coding generates a set of new codes through comparing and merging the codes identified in the open coding phase. ``\textit{Product feedback is not integrated in the pipeline in C3}'' and ``\textit{separation of requirements and development in daily work in C2}'' were further coded as ``\textit{the chasms in DevOps pipelines}''.

After the tentative cores are found from the axial codes, selective coding is conducted to generate the theories following the cores. Around the tentative core ``\textit{the chasms in DevOps pipelines}'', the theory of ``\textit{fragmentary DevOps}'' emerged.

Ethnography studies the practices as phenomena in the natural setting. In the processes of using GT for data analysis, we included some contexts (e.g., \textit{C3} in the label ``\textit{online group communication in C3}'') in the open codes, which might bring some difficulties (e.g., low level of abstraction) to axial coding. However, this could retain the characteristics of natural settings brought by ethnography.
%, where the immersive analysis is often ignored.

%%%\vspace*{-1.0ex}
\subsection{Research Ethics}
Although all the three companies we studied have cooperation agreements with us and approved our research on their development processes, we also take the responsibility of protecting the participating developers' privacy. The research students were insiders of these teams and quietly followed others' normal work style, which helps to protect the privacy of the companies and the developers being observed during the observation process.
%The data on DevOps and microservice practices was collected through participant observations in an environment where the principal investigator and the principal investigators were familiar with the daily activities of the team members. The principal investigator and the principal investigators were insiders of these teams and quietly followed others' normal work style. This helps to protect the privacy of the companies and the developers being observed during the observation process. 
The interviews were conducted with the companies' approvals as well, and all recordings, notes, and documents were handled properly to conform with the confidential agreements.

\section{Findings}
\label{sec:result}
This section presents the findings based on the participant observations and the interviews in this ethnographic study.

\subsection{Organization of Software Teams (RQ1)}
\label{sec:rq1}

\subsubsection{Stubborn organizational structure}
\begin{figure}[!htbp]
  \begin{center}
  \vspace*{-2.0ex}
    \includegraphics[width=0.36\textwidth]{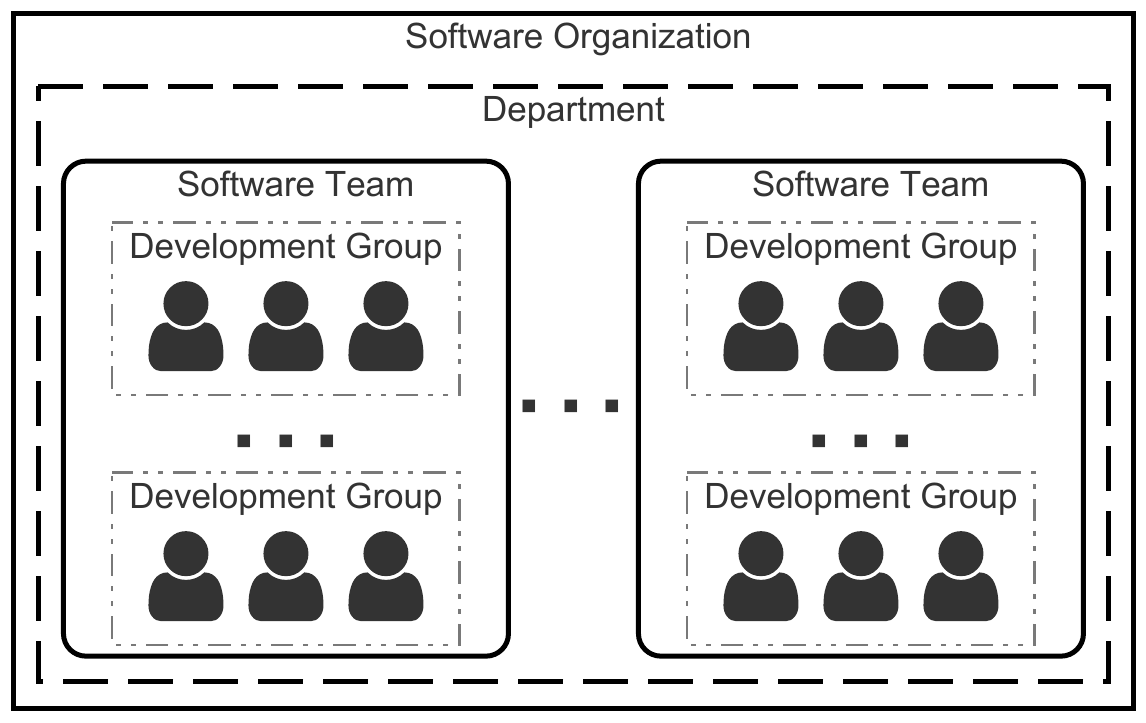}
    \vspace*{-2.0ex}
  \end{center}
    \caption{Three-level organizational structure} 
    \label{fig:structure}
    \vspace*{-2.0ex}
\end{figure}

Although the three companies have different business settings, the organization of their software teams could be similarly abstracted as a three-level structure (Figure~\ref{fig:structure}), which brings challenges to technology-driven changes and culture promotion.

\paragraph{Visionary technology-driven organization} The organizational structure is generally designed and documented by the senior management of the companies, that is, ``\textit{commonly technical-free}''. Other considerations and constraints may come from human resources, finance, and other departments. When the departments attempted internal trials on shifting software teams, the decision makers would consider the ideas and interests of all team members as well as other departments. This process may last a long period and is likely to be rejected, for example in C1, ``\textit{many department heads do not have the gumption to persuade decision makers}''. In Figure~\ref{fig:structure}, technology-driven innovation requires a four-level chain of group-team-department-organization. On the top two levels (department and organization), ``\textit{technology is often not the top concern and option to solve the problem}''.

\begin{figure}[!htbp]
  \begin{center}
  \vspace*{-2.0ex}
    \includegraphics[width=0.46\textwidth]{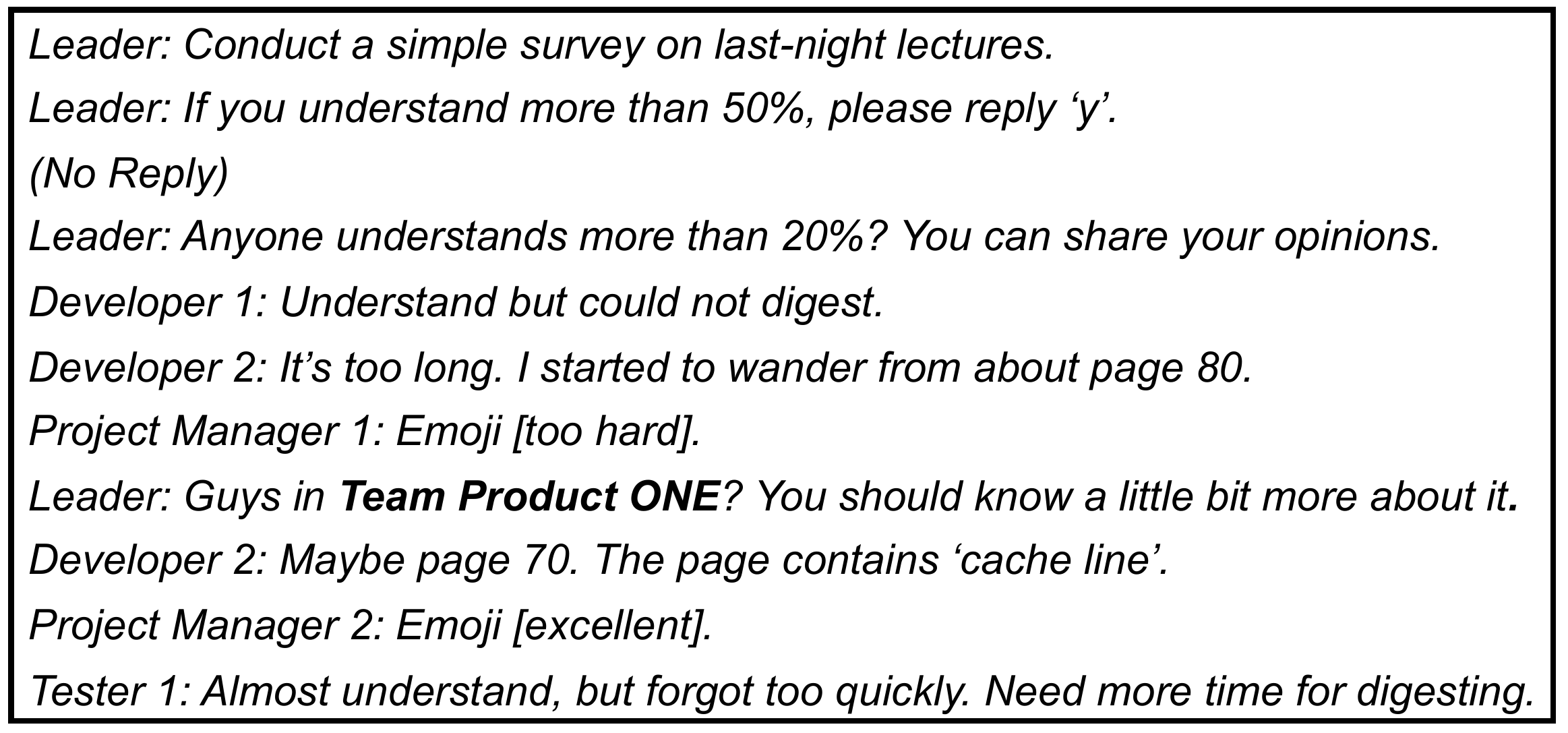}
    \vspace*{-2.0ex}
  \end{center}
    \caption{A conversation after a knowledge sharing in C2} 
    \label{fig:transwarp}
    \vspace*{-2.0ex}
\end{figure}

\paragraph{Deferred culture switch} Trust, the cultural core of DevOps and microservices, needs to be cultivated across software teams. Complaints are still the most prominent collaboration problem among cross-stream collaborations in the three companies. Moreover, some developers are resistant to changes, which would cause extra efforts on them. ``\textit{Developers go into development with a mindset of how they used to develop}'' and the mindset would take a long time to change. Although knowledge-sharing has become a consensus, it takes time to be fostered in software teams. After a knowledge-sharing course on database in C2, merely less than 10\% of the participants expressed their opinion about the content (Figure~\ref{fig:transwarp}), which may indicate that the current climate of knowledge-sharing is not as expected yet.

\subsubsection{Virtual software teams} Affected by the stubborn organizational structure, when using DevOps and microservices for software development, the software team may work as a virtual team.

\begin{figure*}[!htbp]
  \begin{center}
    \includegraphics[width=0.53\textwidth]{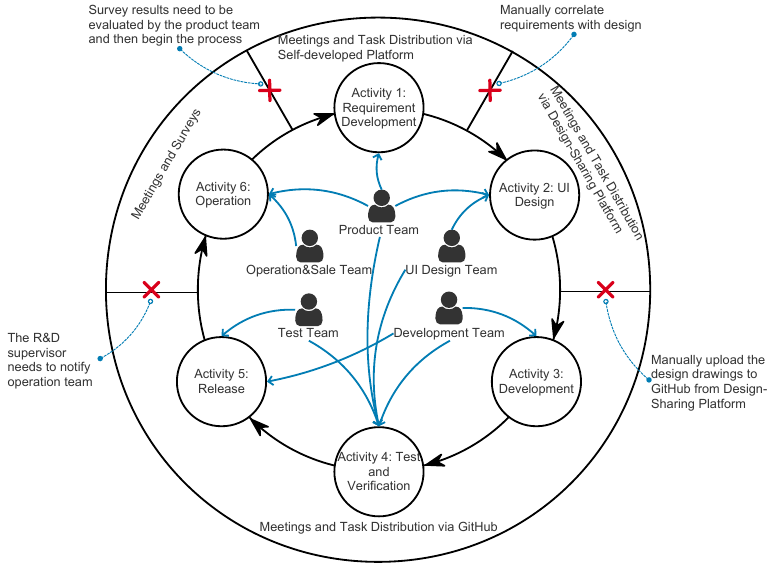}
    \vspace*{-2.0ex}
  \end{center}
    \caption{Observed workflow of a virtual team in C3} 
    \label{fig:SIMU}
    \vspace*{-2.0ex}
\end{figure*}

Figure~\ref{fig:SIMU} shows the workflow of a typical virtual software team for $C3$. Each microservice is maintained by the virtual team composed of engineers, who ``\textit{may take on tasks in different development teams}'', from five teams (i.e. product, UI design, development, test and operations\&sale team), which also ``\textit{achieve the combination of development and operations}''. The development tools for the virtual team were not unified. In $C3$, \textit{self-developed platform} was used for requirement development while \textit{design-sharing platform (Lanhu)} and \textit{GitHub} were used for UI and code management. Similarly, $C2$ uses \textit{JIRA} and \textit{GitLab} respectively for requirements and code management. Virtual teams do not work physical together. They use meetings (e.g., requirements review meeting, code review meeting) to solve problems encountered in understanding microservices and DevOps. An extreme situation experienced was that developers participated in different virtual team meetings throughout the day, leaving no time for development.

\subsection{Gains and Pains of Adopting DevOps and Microservice (RQ2)}

\subsubsection{Gains} Three benefits were found to be significant in observations and were widely mentioned in interviews.

\paragraph{Rapid iteration and deployment} Rapid iteration and deployment are the major benefits from the perspective of the organizations, which allow them to create value to market quickly. Both DevOps and microservices contribute to these benefits. With the adoption of DevOps, ``\textit{the organization sets the rules, builds the basic pipeline, and provides the automatic tools}''. Meanwhile, the adoption of microservices ``\textit{brings new tools and methods, e.g., containers and cloud native to software teams}''. Software teams adopt new techniques and methods to solve problems during implementation of microservices, and the outcomes then will be fed back to the organization. C3, as a startup, employs DevOps and microservices to quickly and consistently deliver on-demand features and release updates once or twice a week. In C2, more than 70\% of teams implement continuous integration on a daily basis, some of them are able to run more than five pipelines a day on average.

\paragraph{Enhanced ability and skills} This can be recognized as the main benefit for individuals that ``\textit{could help them to consider new job opportunities}''. It results from requirements that a single team has to be capable for almost full spectrum of responsibilities in the entire life cycle of DevOps. Another possible reason is that team turnover is likely to be higher in DevOps. Individual's enthusiasm for learning might be stimulated. One interviewee from C1 stated that ``\textit{it is great for engineers to change teams in terms of their development tasks}''.

\paragraph{Reduced burden} This benefit can be attributed to microservices: ``\textit{microservices allow developers to focus on specific domains}''. It further helps ``\textit{increase the happiness of software teams}''. However, the benefit ``enhanced ability and skills'' of DevOps, which requires developers to take on more tasks, may has a negative impact on this.

\subsubsection{Pains} Two major pains while adopting DevOps and microservice were distilled in this study.

\paragraph{High cost} This might be the most significant pain brought by both DevOps and microservices, similar to the problem arising whenever a new technology or method is adopted. C3 took a full year to replace 95\% of the implementations of a particular technology for mobile in the pipeline. 
The consumption of more resources by microservices results in the high cost as well. The shortage of hardware resources, especially for testing, was commonly mentioned in the business meetings in C2. In C1, the cost of deployment constantly increase, multiplying with the number of environments because ``\textit{they have no deployment design}''. Furthermore, locating issues in microservices also leads to high time cost where ``\textit{data consistency needs to be guaranteed while resolving the problems}''.

\paragraph{Lack of practice guidelines} This pain is especially significant in microservices-oriented decomposition. Although many approaches (e.g., Domain-Driven Design and dataflow-based design) for microservices decomposition have been proposed, software teams occasionally felt overwhelmed in their daily work, where the business changes seem out of their control. Along with the business evolution in C1, many microservices began to merge back to larger ones, which become no longer `\emph{micro}' services. C3 showed that when a new business is created, it is difficult and complex to split some new domains or reuse the old domains.

\subsection{Issues of Using DevOps and Microservices (RQ3)}
\label{sec:rq3}

\subsubsection{Fragmentary DevOps}
\label{sec:rq3-1}
Both the above gains and pains are also reported in the previous studies (e.g., \cite{senapathi2018devops, riungu2016devops}), which suggests that DevOps implementations we observed at the three companies to some extent were common to others.
However, some reported benefits such as ``\textit{frequent releases help to reduce stress levels}''~\cite{riungu2016devops} were not directly observed in this ethnographic study, which might be caused by \textit{fragmentary DevOps}.

\begin{figure}[!htbp]
\vspace*{-2.0ex}
  \begin{center}
    \includegraphics[width=0.36\textwidth]{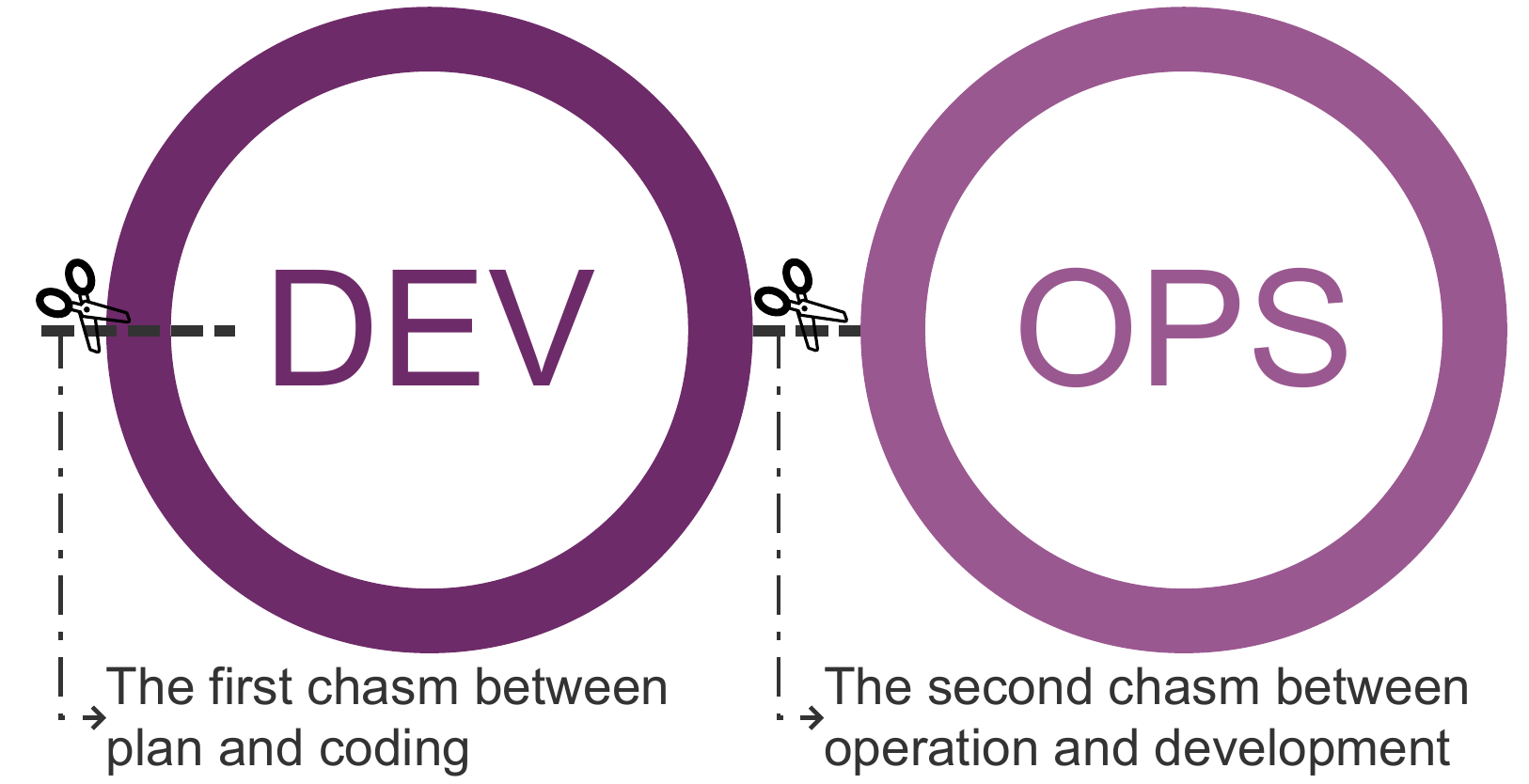}
    \vspace*{-2.0ex}
  \end{center}
    \caption{Two chasms in DevOps pipelines} 
    \label{fig:devops_codes}
    \vspace*{-2.0ex}
\end{figure}

First, software organizations are keen to improve DevOps pipelines (automation, security, etc.), but may have less interest in other essentials (e.g., cross-stream communications) that DevOps advocates for the holistic improvement.
%The developers and software architects were physically separated in C1, which resulted in many details not being communicated in time and further led to some software defects. 
The complete DevOps pipeline was detached into several segments that are barely connected to each other (Figure~\ref{fig:devops_codes}). \textit{The first chasm is between planning and coding.} Although C2 builds an association between JIRA and GitLab, the correlation between requirements and code in the pipeline is elusive. The self-developed project management system in C3 has little to no connection with the code repository. \textit{Another chasm is between operations and others.} For example, operational issues could only be resolved in weekly operations meetings in C3. Moreover, the operations of the Internet infrastructure produced in C1 and C2 are independent from their development. This chasm raises questions about whether DevOps is indeed adopted in organizations because of the separation of `Dev' and `Ops'. Some software teams do not need to use the complete DevOps pipeline in their daily work. Consequently, they only care about the portion of the pipeline that they are involved in. The observed workflow in C3 (Figure~\ref{fig:SIMU}) shows the sense of fragmentation between development activities.

Second, departmental barriers still exist, in some cases deeper than ever. 
%This was observed in C1 and C2, the larger two companies. 
Given the uncertainty of the assignment in some cross-stream tasks, software teams tend to meet their own interests in development. One developer in C1 admitted that ``\textit{he wanted other teams to fork the code instead of asking for more functional requirements}''. The activity of \textit{Test and Verification} in C3 is the most controversial in the entire process because of the involvement of four different teams. Most of the reasons for postponing the release come from this activity.
On the other hand, in the observed three companies, although there exist knowledge-sharing platforms (e.g., public lectures), software teams preferred to share technologies and methods within their own team rather across teams, which might complicate the learning process for other teams. 

All the three companies claim that they have adopted and implemented DevOps. However, the observed adoption of DevOps practices is fragmentary. This calls the need for DevOps maturity, which offers a clearer understanding of the practice levels of software teams and guides the successful adoption of DevOps in software organizations.

\begin{figure}[!htbp]
  \begin{center}
  \vspace*{-2.0ex}
  \includegraphics[width=0.46\textwidth]{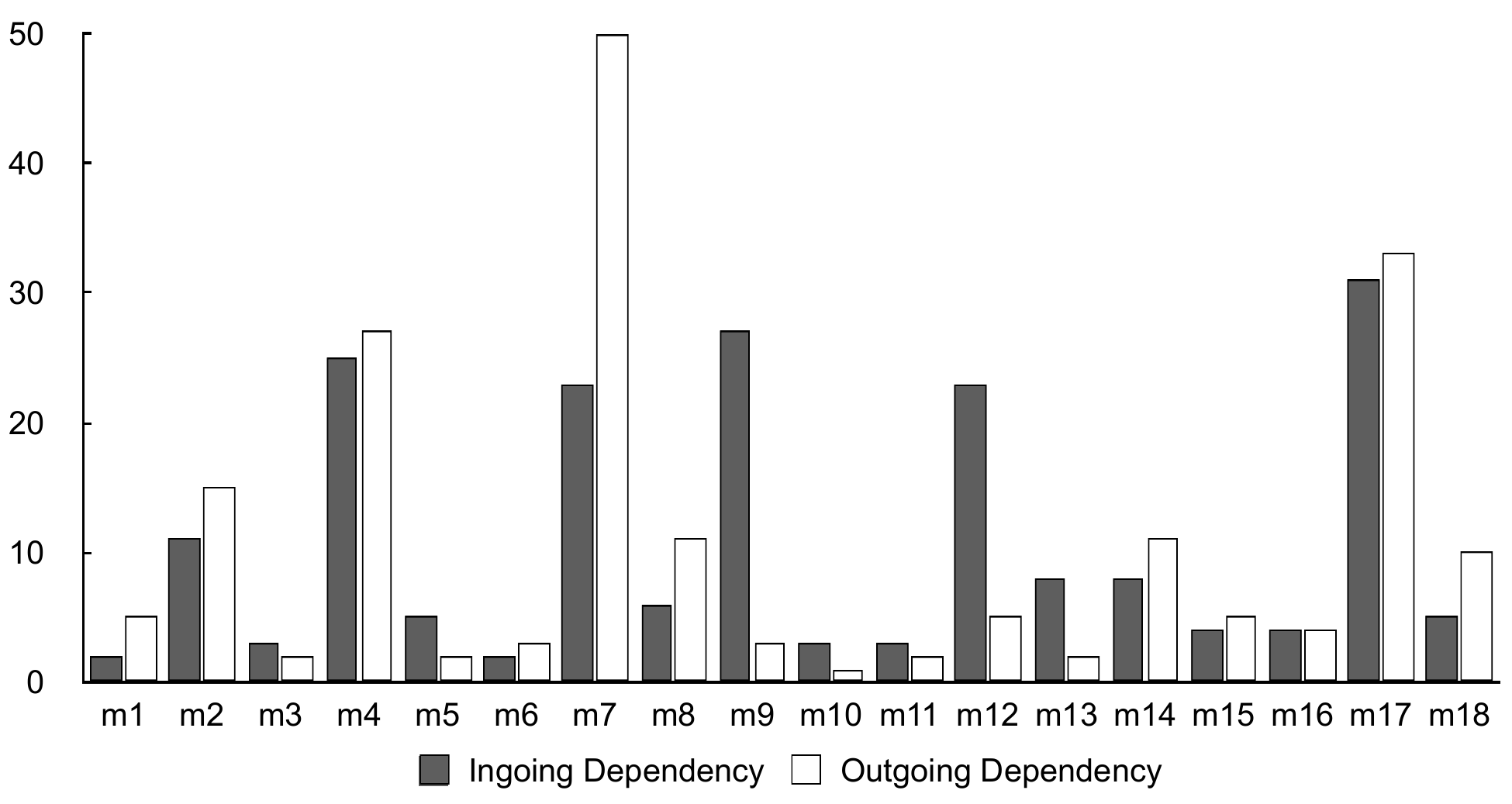}
    \vspace*{-2.0ex}
  \end{center}
    \caption{Dependencies of microservices in C1}
    \label{fig:as}
    \vspace*{-2.0ex}
\end{figure}

\subsubsection{Abuse of microservices}
\label{sec:rq3-2}
When re-thinking of the pains caused by microservices, we found that the \textit{abuse of microservices} might be the major root cause, where ``\textit{blind pursuit of fancy technologies}'' and ``\textit{excessive agility in software development}'' play the important roles. %Figure~\ref{fig:microservices_codes} shows some points of the use of microservices in teamwork. 
%Microservices were introduced to the three companies around 2016, when discussions on microservices rose sharply. 

Microservices may not match all circumstances. The architects in C1 confessed that ``\textit{the implementation of microservices in telecom-context was not a good choice for them at the current stage}''. Figure~\ref{fig:as} shows the dependencies of some microservices in one project in C1. The ingoing and outgoing dependencies of $m4$, $m7$ and $m17$ are significantly overloaded that architects spontaneously merge the mistakenly decomposed microservices. 
Some employees in C2 ``\textit{preferred to SOA rather microservices advocated by the company}''. 

C3 claimed that ``\textit{microservices could now support their business development, but when the requirements accumulate, many problems may arise}''.
%The abuse of microservices is due to the herd mentality, aka ``\textit{blind pursuit of hot technologies}''. 
It might be cool for software organizations or teams to upgrade to new technologies, whereas ignorance of reality, especially business, is likely to be counterproductive. 

\section{Discussion}
This section discusses the relationship between DevOps and microservice in software teams, followed by discussing the lessons learned based on the findings.

\subsection{Relationships between DevOps and Microservices}

\subsubsection{In related work}
\begin{figure}[!htbp]
  \begin{center}
  \vspace*{-2.0ex}
    \includegraphics[width=0.46\textwidth]{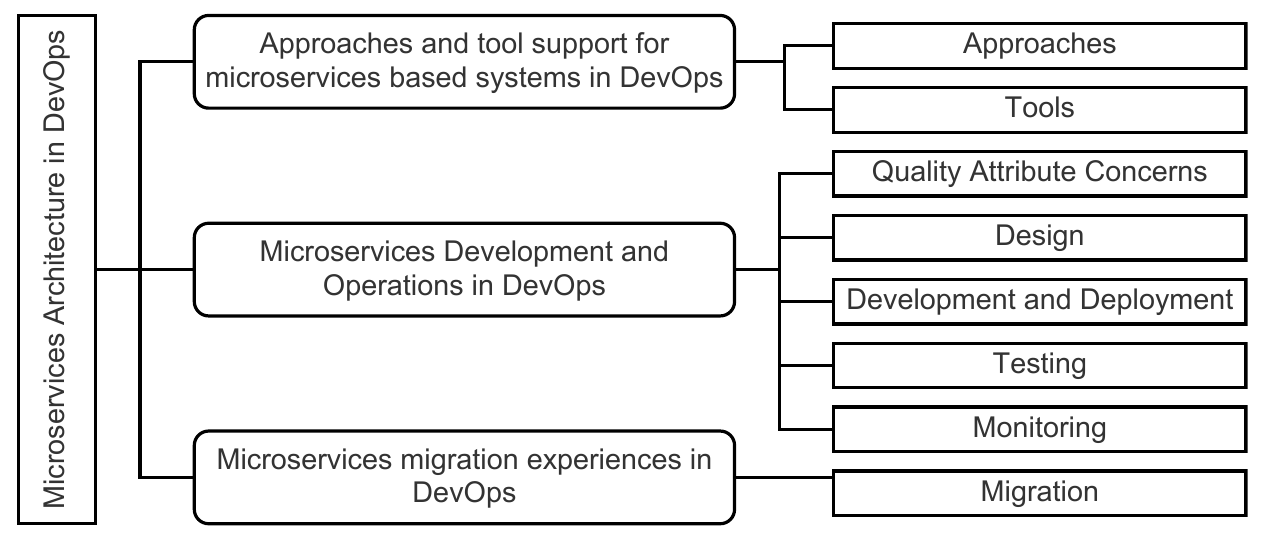}
    \vspace*{-3.0ex}
  \end{center}
    \caption{Classification of research themes on microservices in DevOps~\cite{waseem2020systematic}} 
    \label{fig:microservices_in_devops}
    \vspace*{-2.0ex}
\end{figure}

Researchers considered DevOps and microservices interrelated since their inception.
Enabling effective DevOps implementation by increasing the importance of small teams~\cite{balalaie2016microservices}, microservices are considered as the enabling architecture style for continuous delivery and DevOps~\cite{taibi2018continuous, chen2018microservices}.
Waseem et al. identified 47 primary studies in their mapping study on microservices in DevOps~\cite{waseem2020systematic}, including development and operations of microservices (e.g., \cite{pahl2018architectural}), approaches and tool support for microservices-based systems (e.g., \cite{miglierina2017towards,o2017continuous}), and experiences in migration toward microservices (e.g., \cite{balalaie2015migrating,balalaie2016microservices}), which form the three major themes to describe microservices in DevOps. Figure~\ref{fig:microservices_in_devops} shows the classification summarized in their study. They also collected fifty tools in support of microservices in DevOps. Employing microservices in DevOps was observed to have positive effects on most quality attributes.

\subsubsection{In our observations and interviews}
\begin{figure}[!htbp]
  \begin{center}
   \vspace*{-2.0ex}
    \includegraphics[width=0.48\textwidth]{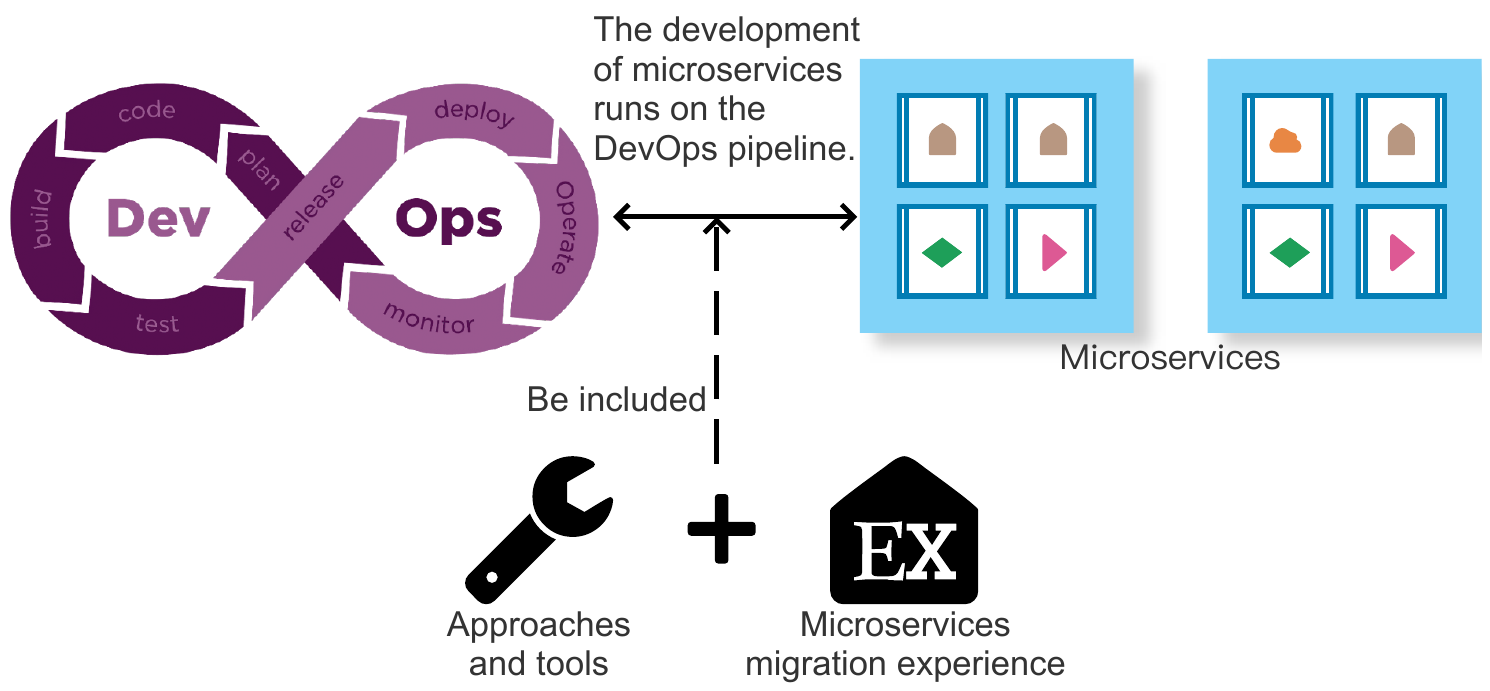}
    \vspace*{-2.0ex}
  \end{center}
    \caption{Relationship between DevOps and microservices} 
    \label{fig:devops_microservices}
    \vspace*{-2.0ex}
\end{figure}

One relationship in~\cite{waseem2020systematic} (the development and operations of microservices in DevOps) can b observed from the three companies in this study, whereas the other two were reflected as a weak relationship (shown in Figure~\ref{fig:devops_microservices}). 

No interviewee voluntarily mentioned the connection between DevOps and microservices. For many developers in the three companies, the connection between the both is limited to ``\textit{microservice systems that they develop were running on the DevOps pipelines}''. The DevOps platform could support the development with or without microservices, meanwhile microservices could be deployed and released in other ways than DevOps pipelines. The software teams and developers are more concerned with getting their daily work done and delivering the project on time than setting up the connection between them, which ``\textit{are most likely to be mentioned in knowledge-sharing sessions}''.

From the perspective of software developers, all the approaches and tools as well as the experiences from migrating to microservices should serve the product development. Consequently, in software teams' daily work, the three aspects reported in ~\cite{waseem2020systematic} could be integrated together as developing microservices on DevOps pipelines. For example, C2 and C3 use Docker to ``\textit{deploy microservices so that the consistency of the development and product environments in DevOps pipelines could be guaranteed to a certain extent}''. 

The lack of joint training on DevOps and microservices could be one reason for the weak correlation. It is observed that although all the three companies offer technical training for their software teams (e.g., cooperative courses with colleges in C1 and `enterprise's college' in C2) and many well-known companies have proposed solutions for microservices with DevOps on Cloud, no courses for the integration of DevOps and microservices in these companies. Moreover, developers are more interested in the technique-specific training courses (e.g., Kafka) and experience courses (e.g., innovative AI solutions for securities industry).

Another significant reason for the disconnection is that opinion leaders lack sufficient knowledge. The department heads or the team leaders, who could be the opinion leaders in software teams, would affect other team members' understanding of such practices as well as the team's development processes. Although they might participate in some courses, their knowledge of the practices is limited. For example, one leader in C3 has recognized that ``\textit{he does not know much about DevOps}''. 
%software developers use more grey literature, where they search for specific technical solutions, than academic literature in their daily work. 
In this case, the participant observation from researchers may trigger the software teams’ awareness of the concepts, thereby improving related practices in teamwork, e.g., security has become a bond between microservices and DevOps in C2. On the other hand, the appeal of researchers is also crucial for the further evolution of practices, e.g., monitoring, security, and performance degradation issues in~\cite{waseem2020systematic}.

% \vspace*{-2.0ex}
\subsection{Lessons Learned}
Section~\ref{sec:result} reports the organization of software teams that adopt DevOps and microservices, as well as the gains, pains, and issues of DevOps and microservices. The lessons for practitioners and the implications for researchers were distilled from four aspects (organization, DevOps, microservices, and their relationship).

% \vspace*{-1.0ex}
\subsubsection{For practitioners}\
% Adopting DevOps and microservice brings benefits and challenges into the software teamwork in the three studied companies. Four lessons were distilled from the observations and interviews for practitioners to improve their DevOps and microservice practices in teamwork.
\vspace*{-1.0ex}
\paragraph{Global perspective} Since the stubborn organizational structure becomes the obstacle against technology-driven changes and culture shift (cf. Section~\ref{sec:rq1}), practice innovation in software teams needs to address the interests on the organization level (NOT single software team) to seek senior management's support to achieve successful implementation.

\vspace*{-1.0ex}
\paragraph{More connections} To tackle the fragmentary DevOps (cf. Section~\ref{sec:rq3-1}), it is essential for software teams to establish the connections between requirements (plan), prototype (design), and code, as well as take the operations on systems into account in development.

\vspace*{-1.0ex}
\paragraph{Active adaption} To mitigate the abuse of microservice (cf. Section~\ref{sec:rq3-2}), practitioners should adapt the architecture to match their domain model, e.g., refactoring microservices and choosing other architectural styles.

\vspace*{-1.0ex}
\paragraph{Joint training} Training on DevOps and microservices together helps practitioners better understand the tasks distributions, thus supports the development in software teams. 

% \vspace*{-1.0ex}
\subsubsection{For researchers} \
% DevOps and microservice are both emerged from Agile developer communities. Thus, it is critical for researchers to review the practical practices in software teamwork. Four implications were also discovered from this ethnographic study.
\vspace*{-1.0ex}
\paragraph{Organization structure} Since the issues of adopting DevOps and microservices come from organization structure (cf. Section~\ref{sec:rq3}), it is worth studying the timing of adjusting organization structure and the methods of avoiding new structures solidification.

\vspace*{-1.0ex}
\paragraph{Maturity model} Although the three companies claimed that they all adopted DevOps, they may have divergent understanding and practice of DevOps, which calls for joint force of academia-industry collaboration to develop the DevOps maturity model. %requires researchers and practitioners to work together to assess the incomplete adoption and implementation of DevOps in software teams.

\vspace*{-1.0ex}
\paragraph{Architecture migration} The migration to microservices has become a popular research topic, where the risk of the abuse of microservices also needs to investigate. Further, the migration to next-generation architecture (e.g., microservices to serverless architecture) is worth exploring by researchers.

\vspace*{-1.0ex}
\paragraph{Practices-oriented courses} The development of a joint training framework (practices-oriented courses) for DevOps and microservices will involve both researchers and practitioners.

\section{Threats to Validity}
\label{sec:validity}
%Some potential threats to the validity in this ethnographic study were identified and tried to mitigate.  

\paragraph{Internal validity}
Reflexivity is very sensitive to ethnologists because of its impacts on the understanding and description of the observed practices~\cite{bibri2019awareness}. The personal experiences as well as social and cultural background may become inherent biases in ethnographic research, which would be reflected in the participant observations. To alleviate this bias, the principal investigator had meetings for discussion in the process of data analysis with other authors who did not participate in all the observations. The raw data by specific engineers (both observation and interview) and codes generated during data analysis are inconvenient to be disclosed in the paper due to corporate confidentiality policies. 
Specifically, the principal investigator describes the observed phenomena and contexts, and then other authors expressed their opinions on these observations from their own perspectives. This constitutes a virtual non-participant observation process. 
Meanwhile, in order to describe the daily work of software teams, the perspectives of the principal investigator and other authors should also be involved in this study. 

\paragraph{External validity}
Although the studied three companies come from different domains and offer rich data on DevOps and microservices in software teams from participant observations and interviews, we do not claim that our results are valid for other scenarios for the reason that these companies share a similar cultural background. Whereas some findings from this study confirm the findings of other studies in different contexts. As this study aims at offering insights into DevOps and microservices in software teams from an immersive perspective, software organizations in similar context could benefit from our findings while other companies could come up with their own findings with reference to our results.

%\section{Threats to Validity}
\vspace*{-1.0ex}
\section{Conclusions}
\label{sec:conclusion}
This paper reports our ethnographic study that combines participant observations and interviews to investigate the adoption and effects of DevOps and microservices in real software teams. We observed three companies from \textit{company culture}, \textit{working environment}, \textit{communications}, \textit{ongoing development processes}, \textit{development environment}, and \textit{meetings}.
%with different type, size, products, customers and degree of globalization from environment, workflow and technology, which could be done with six observation items. 
Eight individual interviews and one group interview were conducted to achieve the data triangulation.

The overall adoption and implementation of DevOps and microservices (e.g., benefits and challenges) by all the three companies is consistent with industry practices reported in previous studies. Since the stubborn organizational structure remains the obstacle against the technology-driven organization and culture switch, virtual software teams are popular in the organizations adopting DevOps and microservices. The three main gains (i.e. rapid iteration and deployment, enhanced ability, and reduced burden) and two major pains (i.e. high cost and lack of practical guidance) are found from the observations and interviews. Furthermore, fragmentary DevOps from the aspects of pipelines and department barriers and abuse of microservices caused by herd mentality and excessive agility are identified as the two issues of DevOps and microservices in software teamwork. We also discuss the weak correlation between DevOps and microservices in software teams. Furthermore, four lessons for practitioners and four implications for researchers are distilled from the findings to implement and study DevOps \& microservices in software teams.

Our ethnographic study contributes to the understanding of software teams that adopt DevOps and microservices from an immersive perspective, which helps understand the interactions between software teams and DevOps \& microservices. We encourage more studies with immersive observations on DevOps and microservices to emerge, so that the adoption and impacts of new technologies and practices in software teams would become traceable, which further enables comparative studies on DevOps and microservices.

\section*{Acknowledgments}

This work is supported by the National Natural Science Foundation of China (No.62072227), the National Key Research and Development Program of China (No.2019YFE0105500) jointly with the Research Council of Norway (No.309494), the Key Research and Development Program of Jiangsu Province (No.BE2021002-2), Intergovernmental Bilateral Innovation Project of Jiangsu Province (No.BZ2020017), as well as the Scientific Research and Innovation Program of Jiangsu Province (KYCX-21-0061).

\vspace*{-1.0ex}
\balance
\bibliographystyle{ACM-Reference-Format}
\bibliography{ICSE}

\end{document}